\documentclass{SciPost}

\hypersetup{
    colorlinks,
    linkcolor={red!50!black},
    citecolor={blue!50!black},
    urlcolor={blue!80!black}
}

\usepackage[bitstream-charter]{mathdesign}
\DeclareSymbolFont{usualmathcal}{OMS}{cmsy}{m}{n}
\DeclareSymbolFontAlphabet{\mathcal}{usualmathcal}
\DeclareMathAlphabet\mathbfcal{OMS}{cmsy}{b}{n}

\usepackage{braket}
\usepackage{bm}
\usepackage{mathtools}
\usepackage{booktabs}
\usepackage{slashed}

\usepackage{framed}
\definecolor{shadecolor}{rgb}{0.95,0.95,0.95}

\newenvironment{asidebox}[1][\hsize]
{%
    \MakeFramed{\hsize#1\advance\hsize-\width\FrameRestore}%
}
{\endMakeFramed}

\fancypagestyle{SPstyle}{
\rhead{{\bf ~ }}

\fancyfoot[C]{\textbf{\thepage}}
}

\begin{document}

\pagestyle{SPstyle}

\begin{center}{\Large \textbf{\color{scipostdeepblue}{
A Lower-Dimensional Remnant of Flux Attachment\\
}}}\end{center}

\begin{center}\textbf{
Gerard Valentí-Rojas\textsuperscript{1,2$\,\star$} and Patrik Öhberg\textsuperscript{2$\,\dagger$}
}\end{center}

\begin{center}
{\bf 1} Naquidis Center, Institut d'Optique Graduate School (IOGS), 91127, Palaiseau, France
\\
{\bf 2} SUPA, Institute of Photonics and Quantum Sciences,Heriot-Watt University, Edinburgh, EH14 4AS, United Kingdom
\\
[\baselineskip]
$\star$ \href{mailto:email1}{\small gerard.valenti-i-rojas@institutoptique.fr}\,,\quad
$\dagger$ \href{mailto:email2}{\small p.ohberg@hw.ac.uk}
\end{center}

\section*{\color{scipostdeepblue}{Abstract}}
\textbf{%
 Flux attachment is a mechanism allowing electric charges to capture magnetic flux in two spatial dimensions. Fundamentally, this is a consequence of the Aharonov-Bohm effect or, in field-theoretic language, of a Chern-Simons term. This is also intimately related to a transmutation of the exchange statistics of the original charges. We show that a remnant of this mechanism is found after a dimensional reduction of a pure Chern-Simons theory and its subsequent coupling to matter.
}

\vspace{\baselineskip}






\vspace{10pt}
\noindent\rule{\textwidth}{1pt}
\tableofcontents
\noindent\rule{\textwidth}{1pt}
\vspace{10pt}



\noindent

\newpage
\section{Introduction}
\label{sec:intro}
Chern-Simons flux attachment lies at the core of the modern panorama of topological phases of matter and anyon physics \cite{fradkin2023field}. Its understanding and widespread use has led to a plethora of applications and a fundamental understanding of quantum Hall fluids \cite{zhang1995chern,fradkin2013,wen2013topological,ezawa2013quantum}. These effects, however, are intimately tied to the dimensionality of the system. One cannot help but wonder whether the characteristic exotic properties are lost or preserved when moving from a system on a spatially two-dimensional surface to another in one dimension. The hope for some preservation of the Chern-Simons phenomenology comes from the fact that fractional statistics are allowed in one spatial dimension \cite{leinaas1977theory}. Elucidating this riddle becomes a pressing issue as modern experiments in ultracold atoms, both in the lattice \cite{kwan2023realization} and continuum \cite{frolian2022realising}, can currently address these phenomena in a controlled environment.

Recent works \cite{valenti2023new,valenti2022topological} argue, contrary to conventional wisdom, that some remnant of a flux attachment mechanism can be found in spatially one-dimensional systems. This comes from the possibility of defining a statistical gauge potential leading to an effective statistical transmutation of matter even when there is no ``flux to attach''. This suggests that the mechanism for the creation of anyons as localised quasiparticle excitations might still remain valid in lower dimensions \cite{valenti2024dual}. Valuable insight can be gained from a controlled reduction of the dimensionality of the problem at hand. Dimensional reduction of matter coupled to a Chern-Simons gauge field was discussed in Refs.\cite{aglietti1996anyons,jackiw2000reduction}. The reduction considered there was deemed trivial in that the dynamics of the gauge field were suppressed along with its exotic phenomenology. However, in the same works, the addition of an external dynamical term ``by hand'' led to a dynamical theory with interesting physical content. 

Here, we provide further evidence in this direction by studying the dimensional reduction of the flux attachment constraint provided by a $\text{U}\,(1)$ Abelian Chern-Simons gauge field.
We show that adequate care in the dimensional reduction yields the aforementioned dynamical contribution in a consistent manner without the need of its introduction ``by hand''. The reduced theory is then coupled to matter fields and shown to induce statistical transmutation and one-dimensional anyonic physics.

It is worth stressing the subtle but fundamentally different approach of other recent studies on the dimensional reduction of anyons. The authors of Refs. \cite{rougerie2023anyons,rougerie2024dimensional} couple a statistical vector potential to matter and then take the thin annulus limit by carefully suppressing a transversal direction. This yields an effective Calogero-type residual interaction in the reduced model and effective elimination of the gauge degrees of freedom. In our approach, we reduce the topological gauge field action and find the reduced analogue of a flux attachment law by considering the coupling to matter \textit{after} the reduction. There is no incompatibility of results, although this signals that the notion of dimensional reduction is vaguely defined, as we indicate throughout the text. 

\subsection{Many-body Aharonov-Bohm Effect}
We start by revisiting the conventional view on flux attachment and statistical transmutation performed by a Chern-Simons gauge field or, equivalently, by a many-body Aharonov-Bohm vector potential. We then give two heuristic arguments for the expectation a dimensional reduction of the flux attachment law to be of a certain functional form. We pursue several strategies for the dimensional reduction of the Chern-Simons term in subsequent sections. 

\begin{figure}[h]
\centering
    \includegraphics[width=0.99\textwidth]{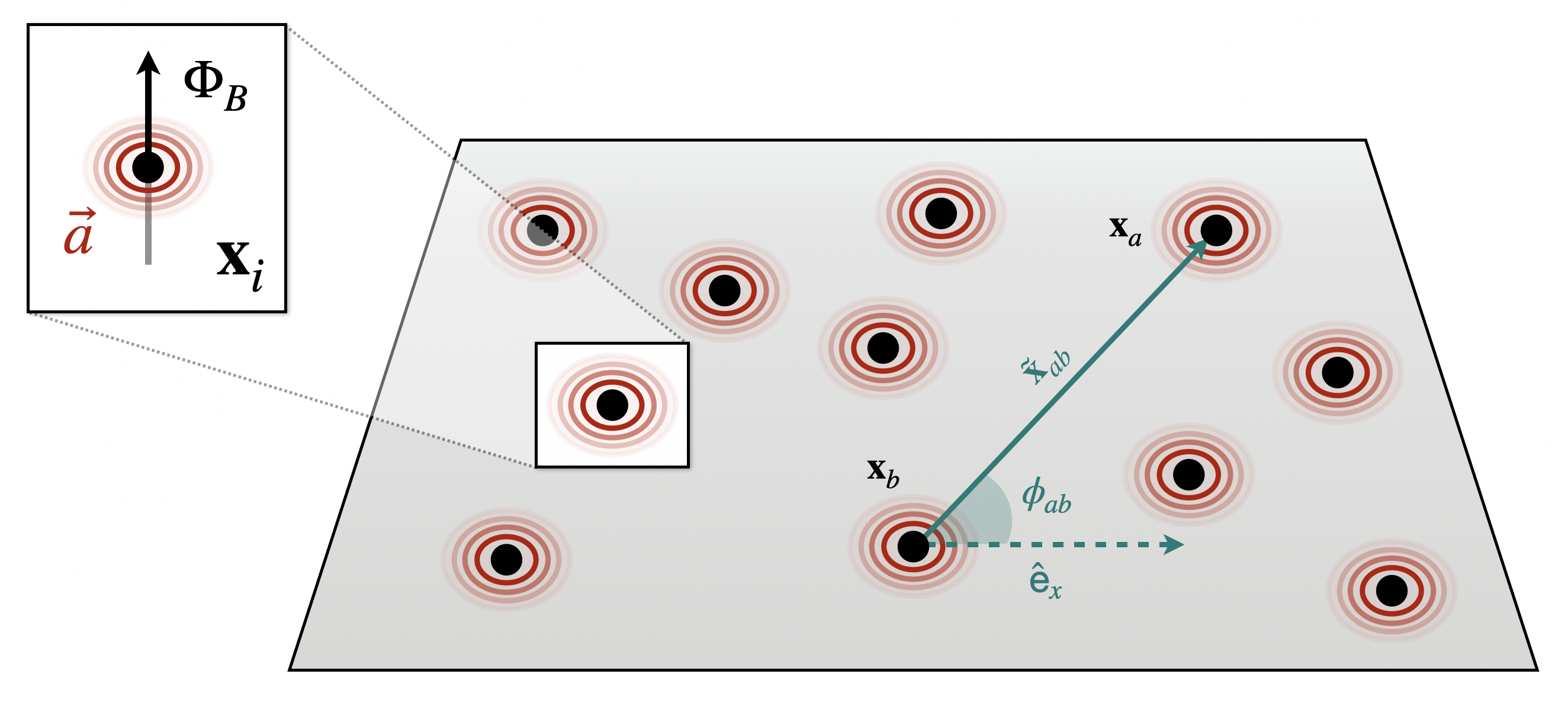}
    \caption{Gas of charge-flux-tube complexes illustrative of a many-body Aharonov-Bohm effect.}
    \label{fig:gas}
\end{figure}

\paragraph{Flux Attachment.} Let us consider a gas of $N$ identical charge--flux-tube complexes, each carrying a magnetic flux $\Phi_{B}$. From a distance, we can think of them as point objects located at position $\mathbf{x}= \mathbf{x}_{i}(t)$. The magnetic field experienced by each object is
\begin{equation}
\bm{b}\, (\mathbf{x}_{i}) = \bm{\nabla}_{\mathbf{x}_{i}}\times \bm{a}\,(\mathbf{x}_{i}) = \sum_{j\neq i} \Phi_{B}\,\delta^{(2)}(\mathbf{x}_{i} - \mathbf{x}_{j})\,\hat{\text{e}}_{z}\;.
\end{equation}
In the Coulomb gauge, the corresponding vector potential is
\begin{equation}
\bm{a}\,(\mathbf{x}_{i}) = \frac{\Phi_{B}}{2\pi} \sum_{j\neq i} \bm{\nabla}_{\mathbf{x}_{i}} \varphi\, (\mathbf{x}_{i} - \mathbf{x}_{j}) = \frac{\Phi_{B}}{2\pi} \sum_{j\neq i} \frac{\hat{\text{e}}_{z} \times (\mathbf{x}_{i} - \mathbf{x}_{j})}{\lvert \mathbf{x}_{i} - \mathbf{x}_{j} \rvert^{2}}\;,
\end{equation}
where $\varphi$ is the polar angle. Defining the number density of point particles as 
\begin{equation}
    n\,(\mathbf{x}) = \sum_{i=1}^{N} \delta^{(2)}(\mathbf{x} - \mathbf{x}_{i})\;,
\end{equation}
we can re-express the initial magnetic field as $\bm{b} \,(\mathbf{x}_{i}) = \Phi_{B}\, n\,(\mathbf{x}_{i})$. This links the magnetic field felt by one particle with the local number density of particles. Such a relation corresponds to a many-body version of the Aharonov-Bohm effect.

\paragraph{Statistical Transmutation.} The natural angular variable is the polar angle, which can take values in $S^{1}$ and has a singularity at $\mathbf{x} = 0$, so the relevant homotopy group is $\pi_{1}(S^{1}) = \mathbb{Z}$, which implies a topological quantisation. This can be formalised by means of the argument function  $\phi_{ab} \equiv \arg\,(\mathbf{\tilde{x}}_{ab}) = \arg\,(\mathbf{x}_{a} - \mathbf{x}_{b}\,; \hat{\text{e}}_{\mathbf{x}})$, where the angle is taken with respect to some arbitrary reference, in this case, the $x$-axis. The exchange property for this function reads $\phi_{ab} = \pm \pi + \phi_{ba}\,$. We can now compute the gauge potential as a non-trivial pure gauge
\begin{flalign}\label{eq:2d_gaugefield}
\bm{a} \,(\mathbf{x}_{i}) &= \alpha\,\nabla_{\mathbf{x}_{i}}  \Phi\,(\mathbf{x}_{1},\dots,\mathbf{x}_{i},\dots,\mathbf{x}_{N}) = \alpha \, \bm{\nabla}_{\mathbf{x}_{i}} \bigg(\sum_{a<b} \phi_{ab}\bigg) = \alpha \, \bm{\nabla}_{\mathbf{x}_{i}} \Big[\sum_{i<b} \phi_{ib} + \sum_{a<i} \phi_{ai} \Big] \\
&= \alpha \, \bm{\nabla}_{\mathbf{x}_{i}} \Big[\sum_{i<b} \phi_{ib} + \sum_{a<i} \big(\pm \pi + \phi_{ia}\big) \Big] = \alpha \, \bm{\nabla}_{\mathbf{x}_{i}}  \Big[\sum_{j\neq i} \phi_{ij} \pm \sum_{a<i} \pi \Big] \\
&= \alpha\, \sum_{j\neq i} \bm{\nabla}_{\mathbf{x}_{i}} \arg\,(\mathbf{x}_{i} -\mathbf{x}_{j}\,; \hat{\text{e}}_{\mathbf{x}}) = \alpha\, \sum_{j\neq i} \bm{\nabla}_{\mathbf{x}_{i}} \arg\,(\tilde{\mathbf{x}}_{ij})\;, 
\end{flalign}
for which one can define a magnetic field and verify that there exists a relation with the charge density $n\,(\mathbf{x}_{i})$ of the form
\begin{equation} \label{eq:local_flux_attach}
	b\,(\mathbf{x}_{i}) = \bm{\nabla}_{\mathbf{x}_{i}} \times \bm{a} \,(\mathbf{x}_{i}) = \alpha \sum_{j\neq i} 2\pi\,\delta^{\,(2)} \,(\mathbf{x}_{i} - \mathbf{x}_{j}) \equiv 2\pi \alpha\, n\,(\mathbf{x}_{i})\,.
\end{equation}
In other words, we naturally recover local flux attachment. Provided the vector potential can be written as a (non-trivial) pure gauge, there is an associated large gauge transformation, in this case is commonly known in literature as a statistical or singular transformation, that removes the gauge field at the expense of altering the statistics under exchange of particles. The transformation reads 
\begin{equation}
	\Psi\,(\mathbf{x}_{1},\dots,\mathbf{x}_{N}) = e^{\,i\alpha \,\sum_{m<l} \text{arg}\, (\mathbf{x}_{m} - \mathbf{x}_{l}\,; \,\hat{\text{e}}_{\mathbf{x}})}\, \Psi_{\text{C}}\,(\mathbf{x}_{1},\dots,\mathbf{x}_{N}) \;,
\end{equation}
where the sum in the exponent is over all particles. Hence, for a given pairwise exchange of two test particles $i \leftrightarrow j\,$, where $1\le i < j \le N$, a corresponding $\pi$ phase from the argument function is collected by the wavefunction for every $e^{-i\alpha \phi_{ab}} e^{i\alpha \phi_{ba}}$ term $i\le a<b\le j$. This yields a statistical factor $\gamma_{ij} = \mp \alpha \pi \eta$, where $\eta \in \mathbb{Z}$ is the number of $a \leftrightarrow b$ possible pairs. This is nothing but a many-particle Aharonov-Bohm phase for flux $\alpha$.

\subsection{Chern-Simons Gauge Theory}
Alternatively, such a peculiar choice of gauge potential is provided by construction, if the correct term is incorporated at the level of a field theoretical Lagrangian. We make use of the quantised Abelian Chern-Simons term at level $1/\alpha$ with $\alpha \in \mathbb{Z}$, minimally coupled to matter via a source term
	\begin{equation}
		S = \frac{1}{4\pi \alpha} \int dt\,d^{2}\mathbf{x}\; \epsilon^{\,\mu \nu \lambda} \hat{a}_{\mu}\partial_{\nu} \hat{a}_{\lambda} - \int dt\,d^{2}\mathbf{x}\; \hat{J}^{\mu} \hat{a}_{\mu}\;.
	\end{equation}
	Computing the Euler-Lagrange equations for the gauge field in the presence of the matter source we are left with
	\begin{equation}
		\hat{J}^{\mu} = \frac{1}{2\pi \alpha} \,\epsilon^{\,\mu \nu \lambda} \partial_{\nu} \hat{a}_{\lambda}\;, \;\;\;\;\;\;\;\;\;\; \partial_{\mu}\, \hat{J}^{\mu} = 0\;,
	\end{equation}
	where the current time component becomes nothing but a constraint equation or Gauss's law of the form
	\begin{equation}\label{eq:gauss_law_app}
		\bm{\nabla} \times \hat{\bm{a}}\,(t,\mathbf{x}) = 2\pi \alpha \, \hat{n}\,(t,\mathbf{x})
	\end{equation}
	which we can attempt to solve in the Coulomb gauge $\bm{\nabla} \cdot \hat{\bm{a}} = 0\,$. This allows us to write the vector potential as  $\hat{\bm{a}} = \bm{\nabla} \times \varphi \,$, so that the Gauss's law becomes 
	\begin{equation}
		\hat{n}\,(t,\mathbf{x}) = \frac{1}{2\pi \alpha} \bm{\nabla}^{2} \varphi\,(t,\mathbf{x})
	\end{equation}
	that can be solved using conventional Green's function methods to find that $\hat{\bm{a}}\,(t,\mathbf{x}) = \alpha \,\bm{\nabla} \,\hat{\Phi}\,(t,\mathbf{x}) $ and
	\begin{equation}
		\hat{\Phi}\,(t,\mathbf{x}) = \int d^{2}\mathbf{x}'\;\varphi\,(t,\mathbf{x} - \mathbf{x}')\,\hat{n}\,(t,\mathbf{x}')\;,
	\end{equation}
	where $\varphi \,(t,\mathbf{x}) = \tan ^{-1}\,(y/x)$ is the conventional polar angle, and where we have assumed that the ``charge'' density is point-like so that the Chern-Simons gauge potential can be written as a pure gauge. This is nothing but the field theoretical version of flux attachment previously found in first-quantised language. The corresponding singular gauge transformation is
 \begin{equation}
	\hat{\Psi}\,(t,\mathbf{x}) = e^{\,i\alpha \,\hat{\Phi} \,(t,\mathbf{x})}\, \hat{\Psi}_{\text{C}}\,(t,\mathbf{x}) \;.
\end{equation}

\section{Dimensional Reduction of Flux Attachment}
The mere possibility of a remnant of flux attachment in 1+1D is rather spectacular. When dynamics happen on a line, there is no Chern-Simons term, and not even a notion of magnetic flux. Yet, there exist Bose-Fermi correspondences \cite{jordan1928paulische,girardeau1960relationship,coleman1975sine,valiente2020long,cheon1999fbduality}, there is potential for anyonic statistics and fractionalisation, and there is anomaly inflow and bulk-edge correspondence. The composite particle duality \cite{valenti2024dual,valenti2022topological} is particularly illuminating in this context, for it tells us that one can still define a statistical gauge field in 1+1D and a composite dual picture, despite all the above. It provides intuition for some results long viewed as obscure, such as the mere definition of anyons in one spatial dimension. The reader, however, might question whether the statistical gauge field in 1+1D has really anything to do with conventional flux attachment. Here, we discuss the dimensional reduction of flux attachment and the emergence of a remnant in one spatial dimension. Let us start by heuristically discussing this through two \textit{gedankenexperiments}. 

\begin{figure}[h]
\centering
    \includegraphics[width=0.99\textwidth]{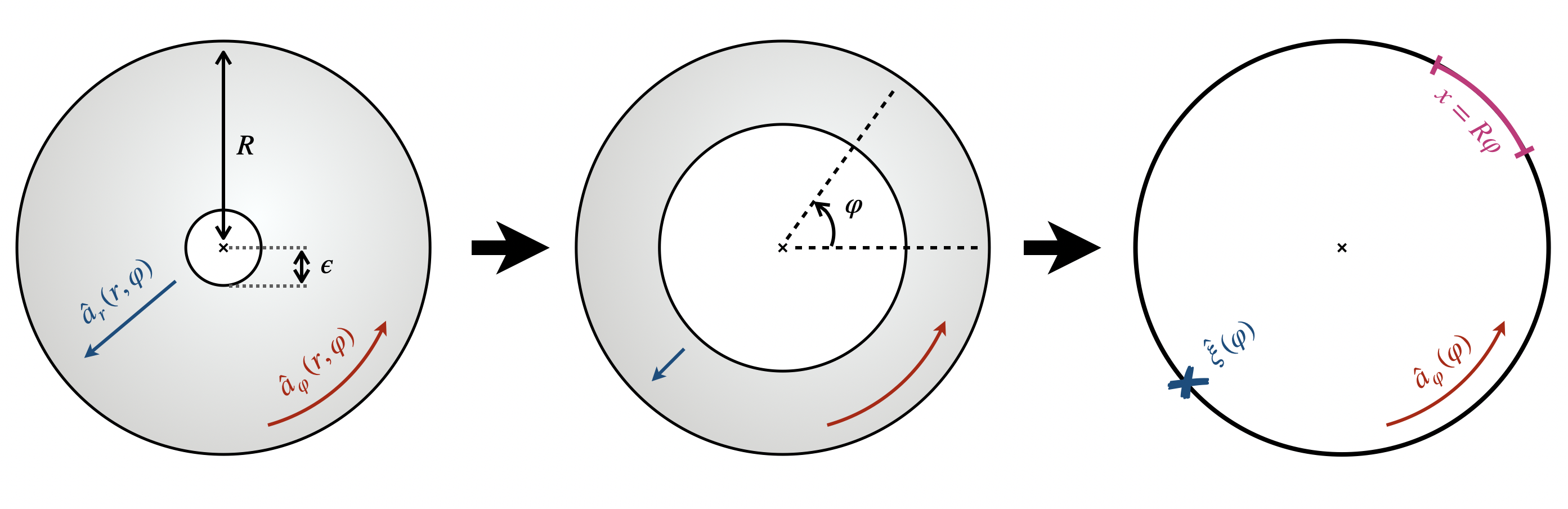}
    \caption{Reduction of a punctured disk to an annulus.}
    \label{fig:punctured}
\end{figure}

\paragraph{A physical Argument.}  Let us consider the local flux attachment law $\bm{\nabla}\times\hat{\bm{a}}\,(\mathbf{x}) = \gamma\, \hat{n} \,(\mathbf{x})$ on a punctured 2d disk with  $r=\epsilon$ inner and $r=R$ outer radius, respectively. This law is assumed to be enforced in the material. When smoothly taking the limit $\epsilon \rightarrow R$ the disk approaches an annulus. This implies $\lvert R-\epsilon\rvert \approx 0$ and $\partial_{r} \,\hat{\bm{a}}\,(r,\varphi)\approx  \mathbf{0}\,$ in $\epsilon \le r \le R$. The magnetic field in polar coordinates $\mathbf{x}=(r,\varphi)$ becomes
\begin{equation}
\bm{\nabla}\times\hat{\bm{a}}\,(\mathbf{x})\,\Big\rvert_{\epsilon \rightarrow R} = \bigg[ \frac{1}{r}\,\hat{a}_{\varphi}(r,\varphi) - \frac{1}{r}\,\partial_{\varphi} \,\hat{a}_{r}(r,\varphi)\bigg] \;\bigg\rvert_{\,r \rightarrow R} \;.
\end{equation}
In this limit, the flux attachment expression effectively decouples from the radial coordinate as it becomes a parameter rather than a spatial dimension (see Figure \ref{fig:punctured}). Thus, at $r=R$, flux attachment reads
\begin{equation}
\hat{\tilde{a}}_{\varphi}(\varphi) = \frac{1}{R}\, \Big[\hat{a}_{\varphi}(\varphi) + \partial_{\varphi} \,\hat{\xi} (\varphi)\Big] = \gamma \,\hat{n} \,(\varphi)\;,
\end{equation}
where $\hat{\xi}(\varphi) = -\hat{a}_{r}(R,\varphi)$ becomes just a memory of the higher dimensional space and is absorbed in a new gauge potential $\hat{\tilde{a}}_{\mu}$. Re-scaling the polar angle as $\varphi \rightarrow x\equiv R\varphi$, we verify that this reduction on an annulus becomes
\begin{equation}
    \hat{\tilde{a}}_{x} (x) = \gamma \,\hat{n}\,(x)\;.
\end{equation}
This is, naively, a possible expression for ``what is left'' of flux attachment on a trivial compact manifold in one spatial dimension. 

\paragraph{A Mathematical Argument.} We have seen that there is a statistical or singular gauge transformation associated to coupling a Chern-Simons gauge field to matter. In fact, the field-theoretical description of Fractional Quantum Hall (FQH) fluids relies on it. It is contained in the form of the statistical gauge potential, typically solved in the Coulomb gauge. We have also seen that such a gauge field has an Aharonov-Bohm-type vortex profile, parametrised by the gradient of the polar angle. We can naturally think about the reduction from the plane to a line by allowing this angle to take only the values $0$ or $\pi$. Thus, effectively we are saying 
\begin{equation}
\varphi (\mathbf{x}) \;\;\;\xrightarrow[\text{reduction}]{\text{dimensional}} \;\;\; \varphi(x) \;\;\;\xrightarrow[\text{to}]{\text{restricted}} \;\;\; \pi\, \Theta (x) = \frac{\pi}{2} \,\big[1+ \text{sgn}\,(x)\big]\;.
\end{equation}
Along with this reduction there is a change in topology, provided the winding around the singularity at $\mathbf{x}= 0$ in 2d, is now only possible if periodic boundary conditions are imposed in 1d. Furthermore, even though the $x=0$ in 1d is ill-defined, it is typically regularised by convention and assumed to be a point of jump discontinuity instead of a singularity. Alternatively, one might choose to use a generalised function $\varepsilon\,(x)$ satisfying Shirokov's algebra \cite{shirokov1979algebra,valiente2021short,valiente2020long}.

Conversely, the $\text{sgn}\,(x)\equiv x/\lvert x \rvert$ for $x\in \mathbb{R}$, defined in one spatial dimension, can be generalised to two dimensions by considering complex coordinates $z=x+i y$ and defining $\text{sgn}\,(z)\equiv z/\lvert z \rvert$ for $z\in \mathbb{C}\,$. But now, the sign function is linked to the polar angle for $z\neq 0$ through the argument function $\text{sgn}\,(z) = \exp{[i\,\arg\,(z)]}$ with the argument function mapping points of the complex plane to the complex unit circle 
\begin{equation}
   \text{arg}\,(z)\,:\; \mathbb{C}-\{\mathbf{0}\} \longrightarrow S^{1}\;,
\end{equation}
but has an associated multivaluedness (or winding) around $z=0$. It is sometimes stated that the singular gauge transformation on the lattice is a 2d Jordan-Wigner transformation. We now see this explicitly provided the disorder operator $\hat{\mathcal{W}}\,(t,\mathbf{x})= \exp{[\frac{i}{\hbar}\,\hat{\Phi}\,(t,\mathbf{x})]} $  for the Chern-Simons theory
\begin{equation}
    \hat{\Phi}\,(t,\mathbf{x}) = \gamma \int d\mathbf{x}'\; \text{arg}\,(\mathbf{x}\,-\mathbf{x}')\,\hat{n}\,(t,\mathbf{x}') 
\end{equation}
reduces to
\begin{equation}
	 \hat{\Phi}\,(t,x) = \pi\gamma \int_{-\infty}^{\infty} dx' \;\,\Theta\,(x - x')\,\hat{n}\,(t,x') = \pi\gamma \int_{-\infty}^{x} dx' \;\hat{n}\,(t,x')\;,
\end{equation}
namely the conventional Jordan-Wigner string. It is worth observing that the string reduces to the usual kink function in when taking the smooth limit of the kernel, in other words $\text{sgn}\,(x) = \lim_{\lambda \rightarrow \infty} \tanh{(\lambda x)}$.

\paragraph{An Important Remark.}A similar logic can be applied in dimensionally reducing the parent topological gauge action. This amounts to reducing the $\text{U} \,(1)$ Abelian Chern-Simons term to 1+1D. This, however, is a hard task for several reasons. Dimensionality is a topological invariant, which means that theories with different dimensionality necessarily belong to different equivalence classes. Secondly, the process of dimensional reduction depends on the protocol used and the topology of the target manifold. The consequence of the former is that it is a one-to-many mapping, meaning that the same parent theory in 2+1D might be consistent with several 1+1D different theories. The Chern-Simons theory has a gauge anomaly in manifolds with boundary, which is precisely the origin of edge states and the bulk-boundary correspondence. It is obscure how these features can be captured in a reduced theory. Finally, the mere notion of \textit{dimensional reduction} is ill-defined or rather ambiguous, provided that a standard Kaluza-Klein reduction on a circle and an edge state on the same geometry can both be considered dimensional reductions of a parent topological gauge theory. All the above signals that such a question is not only an open problem but, formally, a poorly defined one. With that caveat, we shall attempt to find a consistent reduction of the flux attachment theory that provides insights on the link between 2+1D and 1+1D statistical gauge fields. 

\subsection{Dimensional Reduction of an Abelian Chern-Simons Term}

The contraction of the Chern-Simons term with a Levi-Civita symbol and not the usual metric highlights both its topological nature and its dependence on dimensionality. A naïve reduction of the theory is not possible, so elaborate arguments are needed. We can broadly classify the dimensional reduction approaches for the topological gauge term in two types: \textit{(i)} those exploiting the link between bulk and edge degrees of freedom, which effectively integrate the term to the boundary and make use of Stokes' theorem; or alternatively, \textit{(ii)} those in which we can compactify one (or more) dimension(s) and take the limit for which the size of the compact dimension(s) reduce(s) to zero.
\subsubsection{Boundary Theory Reduction}
Let us consider spacetime manifold $\mathcal{M} = \Sigma \times \mathbb{R}$, where $\Sigma$ is a two dimensional spatial manifold with boundary $\partial \Sigma$. A Chern-Simons theory living on such a manifold is not gauge invariant and must be supplied with additional degrees of freedom at the boundary, i.e. edge states. This is known as a gauge anomaly in the bulk, which must be cancelled by the system's edge in order to preserve gauge invariance, and thus, consistency. This is also an illustration of the Callan-Harvey mechanism \cite{callan1985anomalies} for anomaly inflow. Thus, the non-anomalous theory takes into account bulk and boundary degrees of freedom \cite{stone1991edge}, namely
\begin{equation}\label{eq:gauge_var}
S\,[a_{\mu}\,,\xi,\,\dots] = \frac{\kappa}{4\pi} \int_{\mathcal{M}} dt\,d^{2}\mathbf{x}\;\Big[\epsilon^{\,\mu \nu\lambda}a_{\mu}\partial_{\nu}a_{\lambda} + \partial_{\mu}(\xi\,\epsilon^{\,\mu \nu \lambda}\partial_{\nu}a_{\lambda}) \Big] \;\;\;\;+ \;\;\;\;\text{Boundary d.o.f.\;}\;\;\;,
\end{equation}
where we have taken a gauge transformation of the form $a_{\mu} \rightarrow a_{\mu} + \partial_{\mu} \xi\,$. For the particular case in which the manifold is a disk $\Sigma = D_{2}$, the spatial boundary is a circle $\partial \Sigma = S^{1}$, and $\partial \mathcal{M} =S^{1} \times \mathbb{R}$ is a cylinder. The equations of motion reveal that the Chern-Simons gauge connection is flat $f_{\,\mu\nu} = 0$, so the solutions can be written globally as $a_{\mu} = \partial_{\mu}\phi\,$. The Chern-Simons action with gauge variation integrates to the boundary because of the pure gauge nature of solutions, yielding
\begin{equation}\label{eq:bdy_action}
	S_{\text{CS}}[a_{\mu}\,,\xi]\longrightarrow S_{\text{\,bdry}} \,[\phi_{L,R},\xi] = \frac{\kappa}{4\pi} \int_{\,\partial \mathcal{M}} dt\,dx\;\big(\pm 
 \partial_{t}\,\phi_{L,R}\,\partial_{x}\,\phi_{L,R} + \phi_{L,R}\,\epsilon^{\,\mu \nu}\partial_{\mu}a_{\nu} \big)\;,
\end{equation}
which is the action of a chiral scalar field with an axion-like term that appears as a consequence of the gauge anomaly. The $\pm$ sign together with $L,R$ subindices are meant to capture the two chiralities that may appear, e.g. if we choose an annulus there are counter-propagating chiral bosons, on  each edge. The holographic connection between bulk Chern-Simons theory and boundary Rational Conformal Field Theories (R-CFTs) was initially found by Witten \cite{witten1989quantum} and Moore-Seiberg \cite{moore1989taming}. This can be considered as a dimensional reduction provided final theory lives in 1+1D, while the bulk one lives in 2+1D. A purely geometric decomposition (see Appendix \ref{sec:divergence}) is consistent with this approach. 

\paragraph{Revisiting Wen's Edge State Action.}
Wen \cite{wen1990chiral,wen1990topological,wen1991topological} used the aforementioned approach to find the edge state theory of a fractional quantum Hall fluid. In order to see this, let us start from the Chern-Simons action in the temporal gauge
\begin{equation}
\mathcal{L}_{\text{CS}} \,[a_{0} = 0] = -\frac{\kappa}{4\pi} \epsilon^{\,0ij} a_{i} \,\partial_{0}\, a_{j}\;.
\end{equation}
Substituting the pure gauge solution $a_{i}  = \partial_{i}\,\phi$ back in the Chern-Simons action in the $a_{0} = 0$ gauge
\begin{equation}
\begin{split}
S_{\text{CS}}\, [a_{i}] &= -\frac{\kappa}{4\pi} \int dt d^{2}\mathbf{x}\; \epsilon^{\,ij} a_{i} \,\partial_{0}\, a_{j} = -\frac{\kappa}{4\pi} \int dt d^{2}\mathbf{x}\; \Big[ \partial_{x}\,(\phi\,\partial_{t} \,\partial_{y}\,\phi )- \partial_{y}\,(\phi\,\partial_{t} \,\partial_{x}\,\phi )\Big] \\
&= -\frac{\kappa}{4\pi} \int dt d^{2}\mathbf{x} \;\big(\bm{\nabla} \times \mathbf{v}\big)_{z} = -\frac{\kappa}{4\pi} \int dt \oint_{C} d\mathbf{l}\cdot \mathbf{v}\;,
\end{split}
\end{equation}
where we have used Stokes' theorem to reduce dimensionality in the last step, and previously defined $\epsilon^{\,ij} \equiv \epsilon^{\,0ij}$ and $\mathbf{v} = (\phi\,\partial_{t} \,\partial_{x}\,\phi, \phi\,\partial_{t} \,\partial_{y}\,\phi)\,$. Hence, we obtain a boundary action
\begin{equation}
S_{\mathrm{edge}}\,[\phi] = \frac{\kappa}{4\pi} \int_{\partial \mathcal{D}} dt\,dx\; \phi\,\partial_{t} \,\partial_{x}\,\phi = \frac{\kappa}{4\pi} \int dt\; \bigg\{  \Big[ \phi\,\partial_{x}\,\phi\Big]\Big|_{x=0}^{x=2\pi} - \int  dx\;\partial_{t} \, \phi \,\partial_{x} \, \phi \bigg\}\;.
\end{equation}
Notice that the last term is a chiral boson constraint, and it is purely topological, meaning that it is independent of the particular system we consider and will have $H_{\mathrm{edge}} = 0$. There is no kinetic term in the Hamiltonian, so no sense of velocity for edge state. Wen \cite{wen1990chiral,wen1991topological,wen2013topological} fixed this by adding a non-topological contribution that depends on the system, to have a velocity dependent on the confining potential. This can be incorporated in a natural way by a mere change of coordinates. Such a contribution does not come from the Chern-Simons term, so we avoid its inclusion here but we refer the interested reader to the original works \cite{wen1990chiral,wen1991topological,wen2013topological}.

\paragraph{Coupling to a Background.} Let us now minimally-couple the edge chiral boson to a background electric field \cite{levkivskyi2012theory}. The minimal coupling for a chiral boson to the background gauge field $A_{\mu}$ is given by the gauge-covariant derivative, defined as $D_{\mu} \phi \equiv \partial_{\mu} \phi - q A_{\mu}\,$, and the boundary Lagrangian density becomes
\begin{equation}
\mathcal{L}_{\text{g-edge}}\,[A_{\mu}\,,\phi]=\pm \frac{\kappa}{4\pi} D_{t}\phi D_{x}\phi + \gamma\, \epsilon^{\,\mu \nu}\, \partial_{\mu}\,\phi \,A_{\nu}
\end{equation}
with $\mu = \{0,1 \} = \{t, x \}$ now being the coordinates of this effective 1+1D theory, which constitutes a model for a $\text{U} \,(1)$--gauged chiral boson. We can now perform the gauge transformations 
\begin{equation}\label{eq:gauge_transf}
\phi \longrightarrow \phi + q \xi \;\;\;\;\;\;\;\;\;\; \text{and} \;\;\;\;\;\;\;\;\;\; A_{\mu} \longrightarrow A_{\mu} + \partial_{\mu}\,\xi\;.
\end{equation}
The first term is invariant under gauge transformations. On the other hand, the second term transforms like 
\begin{equation}\label{eq:variation_gauge}
\delta(\gamma\, \epsilon^{\,\mu \nu}\, \partial_{\mu}\, \phi \, A_{\nu}) = \gamma\, \epsilon^{\,\mu \nu}\, \partial_{\mu}\,\phi \,\partial_{\nu}\,\xi +q\gamma\,\epsilon^{\,\mu \nu}\, \partial_{\mu}\,\xi \,\partial_{\nu}\,\xi  + q \gamma\, \epsilon^{\,\mu \nu} \,\partial_{\mu}\,\xi \,A_{\nu} \;.
\end{equation}
The first contribution in Eq. \eqref{eq:variation_gauge} vanishes upon integration by parts, the second is explicitly zero provided $\xi$ is non-singular in the $t-x$ plane, and only the third one survives integration, yielding
\begin{equation}
\delta \mathcal{L} = - q \gamma\,\xi (t,x)\,\epsilon^{\,\mu \nu}\, \partial_{\mu}\,A_{\nu}(t,x)\;.
\end{equation}
This shows that the gauged chiral boson has a gauge anomaly of the same form as the gauge variation in  the bulk Chern-Simons theory in Eq. \eqref{eq:gauge_var}, which is crucial to cancel the gauge anomaly in the FQHE. Hence, the mixed-dimensional action \eqref{eq:gauge_var} can be rewritten explicitly as \cite{stone1991edge}
\begin{equation}
    S[a_{\mu}\,,\phi] = S_{\text{CS}}[a_{\mu}] + S_{\text{g-edge}}[a_{\tilde{\mu}}\,,\phi]\;.
\end{equation}
This is now gauge invariant upon transformations of the form of Eq. \eqref{eq:gauge_transf}. Notice that $\mu = 0,1,2$ while $\tilde{\mu}=0,1$.
\paragraph{Boundary Reduction.} We consider our topological gauge action to be
\begin{equation}
    S\,[A_{\mu}\,,a_{\mu}] = \frac{1}{4\pi} \int_{\mathcal{M}} dt\,d^{2}\mathbf{x}\;\Big[\epsilon^{\,\mu \nu\lambda} A_{\mu}\partial_{\nu}\,\big(\kappa A_{\lambda} - 2 a_{\lambda} \big)\Big]\;.
\end{equation}
We can integrate out one of the gauge fields and express it as a normal Chern-Simons term in terms of the remaining statistical gauge field. According to the previous discussion, this model has a boundary theory
\begin{equation}\label{eq:reduced}
S_{\,\text{bound.}} \,[\phi_{L,R}\,,a_{\mu}] = \,\frac{1}{4\pi} \int_{\partial \mathcal{M}} dt\,dx\;\Big(\pm \kappa\,\partial_{t}\,\phi_{L,R}\,\partial_{x}\,\phi_{L,R} -  \phi_{L,R} \,\epsilon^{\,\mu \nu}\,\partial_{\,\mu}\,a_{\nu} \Big)\;.
\end{equation}
Thus, considered as a dimensional reduction, we can write the low-dimensional encoding as $S_{2+1} \rightarrow S_{1+1}$, with identification
\begin{equation}
   S_{\text{CS}}\,[A] +S_{\text{BF}}\,[A, a] \;\;\;\;\;\longrightarrow \;\;\;\;\; S_{\chi}\,[\phi] + S_{\text{axion}}\,[\phi, a] \;.
\end{equation}
This reduced action constitutes the 1+1D equivalent of the flux attachment action in 2+1D. It is worth noting that the reduced theory is manifestly anomalous in that the chiral boson satisfies a $\text{U} \,(1)$ Kac-Moody algebra. As predicted by the composite particle duality \cite{valenti2023new,valenti2022topological}, non-trivial consequences are expected when such a gauge theory is coupled to matter. Hence, the field theory of a topological fluid in 1+1D is
\begin{equation}
   S = S_{\,\chi-\text{axion}} \,[\phi_{L,R}\,,a_{\mu}] +S_{\text{matter}}\,[\bar{\Psi},\Psi] - \int dt dx \; J^{\mu}a_{\mu}  \;.
\end{equation}


\noindent The above discussion does not assume or make any reference to a coupling to a specific type of matter, it is purely based on topological gauge theory arguments. It is worth comparing such a model with the usual edge state theory of FQH fluids, given by
\begin{equation}
    S^{\text{FQH}}_{\text{edge}}\,[\phi_{L,R}\,,A_{\mu}] = \frac{\kappa}{4\pi}\int_{\partial \mathcal{M}} dt dx\;\Big(\pm  D_{t}\phi_{L,R} D_{x}\phi_{L,R} - \mathcal{H}(\rho) + \epsilon^{\,\mu \nu}\, \partial_{\mu}\,\phi_{L,R}\,A_{\nu} \Big)
\end{equation}
where $\mathcal{H}$ is the Hamiltonian density in terms of the matter density $\rho = \pm \frac{\kappa}{2\pi} D_{x}\phi_{L,R}$ characterising a deformation at the edge, i.e. a chiral density wave. Notice that the last term, which provides the coupling to a background gauge field, is also known as a many-body Aharonov-Bohm twist. This means that for edge states in a FQH liquid, the chiral boson $\phi_{L,R}$ acquires the meaning of a field characterising the boundary density fluctuations.  In a linear approximation, the edge state is considered to be described by $\mathcal{H}(\rho) = v (D_{x} \phi_{L,R})^{2}$, where $v$ is the group velocity of the edge excitation. In the linear regime and for $A_{\mu} = 0$, one recovers Wen's chiral Luttinger liquid \cite{wen1990chiral}. Nonetheless, one can include additional terms in $\mathcal{H}(\rho)$ that dominate the dynamics at long times. This yields nonlinear descriptions of FQH edge states \cite{bettelheim2006,wiegmann2012,nardin2023nonlinear}.

 \subsubsection{Compactified Theory Reduction}

\paragraph{Temporal Gauge.} Let us now consider the more traditional approach to dimensional reduction based on compactification of dimensions. Provided a gauge theory living in 2+1D Minkowski space, we take one spatial dimension and impose periodic boundary conditions. More explicitly, given a gauge field living in the spatial $x-y$ plane with length $L_{x}\times L_{y}$, we compactify the $y$-direction into a circle of radius $R =\frac{1}{2\pi} L_{y}$, so that the spatial manifold is now parametrised as a hollow cylinder of length $L_{z} = L_{x}$ and radius $R$. A gauge field living on this manifold has the form $a_{\mu}(x^{\mu})\equiv a_{\mu} (t,R,\varphi,z) = (a_{t},a_{\varphi},a_{z})$. The magnetic flux penetrating a closed contour $\mathcal{C} = S^{1}$ that we choose as circular, with radius $r=R$ and oriented along the positive $z$-axis of the cylinder, is
\begin{equation}
    \Phi(t,z) = \int_{D_{2}} d\mathbf{S}\cdot(\bm{\nabla} \times\bm{a}) = \int_{S^{1}} d\mathbf{l}\cdot \bm{a} = R \,\oint_{S^{1}}d\varphi\;a_{\varphi}(t,R,\varphi,z)\;.
\end{equation}
We can now imagine taking the thin-cylinder limit $R\ll 1$, so that the gauge dynamics along $\varphi$ are effectively suppressed. This motivates the change $a_{z}(t,R,\varphi,z) \rightarrow a_{z}(t,z)$. Then, the Chern-Simons action in the temporal gauge reads
\begin{flalign}
S_{\text{CS}} [a_{0} = 0] &=\frac{\kappa}{4\pi} \int dt d^{2}\mathbf{x} \; \epsilon^{\,i0j} a_{i} \,\partial_{0}\, a_{j} \\
&= \frac{\kappa}{4\pi} \int dt dz \;\Big[ -a_{z} \partial_{t} \big(R\oint d\varphi \,a_{\varphi}\big) + \big(R\oint d\varphi \,a_{\varphi}\big) \partial_{t}a_{z}\Big] \\
&= \frac{\kappa}{2\pi} \int dt dz \; \Phi (t,z) \,\partial_{t} a_{z} (t,z) \;\;\; + \;\;\text{Boundary Term}\;.
\end{flalign}
So, the Chern-Simons action reduces to a locally-varying $\theta$-term, or 1+1D axion term, or polarisation term, or Background Field (BF) term $\propto \Phi \hat{E}_{z}$. Notice that defining a holonomy in the reduced transverse dimension amounts to calculating the Wilson loop, which is invariant under gauge transformations
\begin{equation}
    \hat{\mathcal{W}}\,[\Gamma] = \exp{\bigg(\frac{i}{\hbar}\,\oint_{\Gamma}dx^{\varphi}\,a_{\varphi} + \partial_{\varphi}\xi\bigg)}  = \exp{\bigg(\frac{i}{\hbar} \,\Big[\Phi(t,z) + 2\pi k \Big] \bigg)}
\end{equation}
for $k\in \mathbb{Z}$, implying that the magnetic flux, now a scalar field in 1+1D, has to be compact $\braket{\Phi}\in [0,2\pi)$ --- in essence an angle or phase variable --- in order to preserve gauge invariance, meaning $\Phi \rightarrow \Phi + 2\pi k$. 

\begin{asidebox}
\noindent
\textbf{Geometrical Protocol for Dimensional Reduction}\\

\noindent
The dimensional reduction via compactification of a spatial direction on a circle can be geometrically summarised as
 \begin{equation}
M^{3} \simeq \mathbb{R}^{1,2}\;\;\;\xrightarrow[\text{of space}]{\text{compactification}} \;\;\; M^{2}\times S^{1} \simeq \mathbb{R}^{1,1}\times S^{1}  \;\;\;\xrightarrow[\text{to}]{\text{reduction}} \;\;\; M^{2}\simeq \mathbb{R}^{1,1}\;,
\end{equation}  
where $M^{D}$ denotes a $D$-dimensional Minkowski spacetime.
\end{asidebox}

\begin{figure}[h]
\centering
    \includegraphics[width=0.99\textwidth]{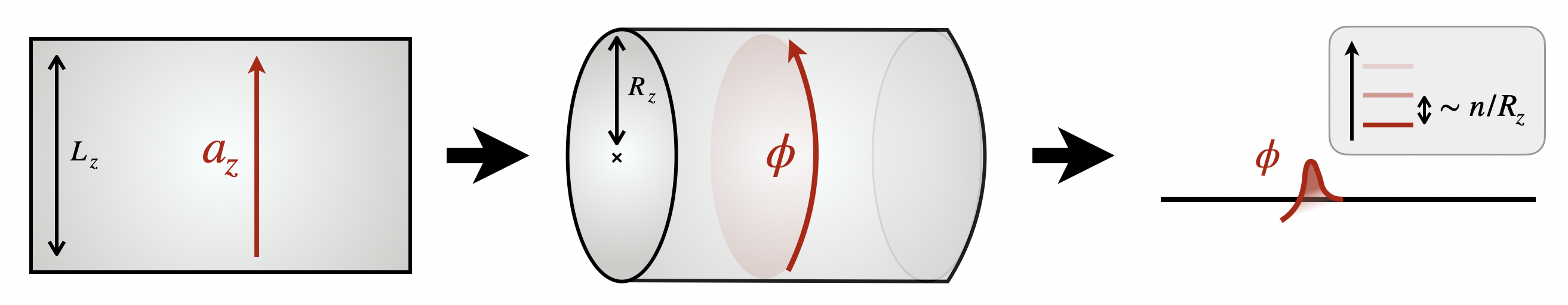}
    \caption{Compactification on a circle and Kaluza-Klein reduction.}
    \label{fig:kkred}
\end{figure}

\paragraph{Covariant Kaluza-Klein.}The previous argument can be incorporated within the conventional Kaluza-Klein prescription for dimensional reduction, which allows an explicitly covariant treatment. We consider the particular circular compactification $\mathcal{M}_{D} = \mathcal{M}_{D-1}\times S^{1}$ and coordinates $\tilde{x}^{\,\mu} = (x^{\,\mu}, z)\,$, where $z$ labels the compact dimension, with length $L_{z} = 2\pi R_{z}\,$. Fields must satisfy a periodicity condition of the form $\Phi\,(x^{\,\mu}, z + 2\pi R_{z}) = \Phi\,(x^{\,\mu}, z)$ which allows an expansion in terms of Fourier --- also known as Kaluza-Klein --- modes. For a scalar field, this is
\begin{equation}
	\Phi\,(x^{\,\mu}, z) = \sum_{n=-\infty}^{\infty} \Phi_{n}\,(x^{\,\mu})\,e^{\,in\frac{z}{R_{z}}}\;.
\end{equation}
These modes are gapped with a mass that scales like $m \sim n/R_{z}$. In the $R_{z} \rightarrow 0$ limit, the mass gap grows and the relevant modes are those close to $n = 0$. Hence, in this limit
\begin{equation}
    \Phi (x^{\,\mu},z) \approx \Phi_{0}(x^{\,\mu}) \equiv \tilde{\Phi}(x^{\,\mu})\,.
\end{equation}
For a gauge field $a_{\,\nu} \,(x^{\,\mu},z) = \big(a_{\bar{\nu}}\,(x^{\,\mu},z), a_{z}\,(x^{\,\mu},z)\big)$ the story is similar but we have to distinguish between components perpendicular to the compact direction and those pointing along the internal direction
\begin{flalign}
	&a_{\,\bar{\nu}}\,(x^{\,\mu}, z) = \sum_{n=-\infty}^{\infty} a_{\,\bar{\nu}}^{\,(n)}\,(x^{\,\mu})\,e^{\,in\frac{z}{R_{z}}} \;\;\;\xrightarrow[\text{limit}]{\text{thin cylinder}} \;\;\; a_{\,\bar{\nu}}\,(x^{\,\mu}, z) \approx a_{\,\bar{\nu}}^{(0)} (x^{\,\mu})\equiv \tilde{a}_{\,\bar{\nu}}(x^{\,\mu}) \;,\\
    &a_{\,z}\,(x^{\,\mu},z) = \sum_{n=-\infty}^{\infty} a_{\,z}^{\,(n)}\,(x^{\,\mu})\,e^{\,in\frac{z}{R_{z}}} \equiv \phi\,(x^{\,\mu},z)\;.
\end{flalign}
Upon integration over the compact dimension, the latter becomes 
\begin{equation}
     \oint_{S^{1}} dz\;\phi \,(x^{\,\mu},z) \equiv \tilde{\phi} (x^{\mu}) \in [0,2\pi)\;,
\end{equation}
which can be interpreted as the flux through the cylinder at a given reduced spacetime position. Applying this scheme to the Abelian Chern-Simons term, we might verify
\begin{flalign}\label{eq:kkcs}
	& \text{lim}_{\,R_{z} \rightarrow \,0} \,\bigg[\frac{\kappa}{4\pi} \int _{\,\mathcal{M}\,\times\, S_{1}}  d\tilde{x}^{\,\mu}\;\epsilon^{\,\mu \nu \lambda}\,a_{\mu}(\tilde{x}^{\,\mu}) \,\partial_{\nu}\,a_{\lambda}(\tilde{x}^{\,\mu})  \bigg]\\
 =& \frac{\kappa}{4\pi} \int_{\,\mathcal{M}} dt\, dx\; \Big( 2 \oint_{S^{1}}dz\;a_{z}(x^{\,\mu},z)\Big)\,\epsilon^{\,z \bar{\nu} \bar{\lambda}} \partial_{\bar{\nu}}\, a^{\,(0)}_{\bar{\lambda}}(x^{\,\mu})\\ \label{eq:theta_reduced}
=& \frac{\kappa}{2\pi} \int_{\,\mathcal{M}} dx^{\,\mu}\;\tilde{\phi}(x^{\,\mu})\,\epsilon^{\bar{\nu} \bar{\lambda}} \,\partial_{\bar{\nu}}\, \tilde{a}_{\bar{\lambda}}(x^{\,\mu}) \;.
\end{flalign}
So we have effectively reduced the theory from $D$ to $D-1$ dimensions, for $D=2+1$. See that $\mu$ correspond to the indices for the parent $D$-dimensional theory, while $\bar{\mu}$ refer to the reduced theory in $D-1$ dimensions. Observe that the gauge variation on the parent Chern-Simons action can also be reduced by the same protocol. For the gauge transformation $a_{\mu} \rightarrow a_{\mu} + \partial_{\mu}\xi\,$, this results in
\begin{flalign}
   \text{lim}_{\,R_{z} \rightarrow \,0} \,[ \delta S_{\text{CS}} ] =& \;\text{lim}_{\,R_{z} \rightarrow \,0} \,\bigg[\frac{\kappa}{4\pi} \int _{\,\mathcal{M}\,\times\, S_{1}}  d\tilde{x}^{\,\mu}\;\epsilon^{\,\mu \nu \lambda}\,\partial_{\mu}\, \xi(\tilde{x}^{\,\mu}) \,\partial_{\nu}\,a_{\lambda}(\tilde{x}^{\,\mu})  \bigg] \\
    =&\; \frac{\kappa}{4\pi} \int_{\,\mathcal{M}} dx^{\,\mu}\;\epsilon^{\bar{\mu} \bar{\nu}}\partial_{\bar{\mu}}\,\tilde{\xi}(x^{\,\mu})\,\partial_{\bar{\nu}}\, \tilde{\phi}(x^{\,\mu}) \\\label{eq:twisted_reduced}
    =&\; \frac{\kappa}{4\pi} \int_{\,\mathcal{M}} dt\,dx\; \Big( \partial_{t}\,\tilde{\xi}\,\partial_{x}\, \tilde{\phi} - \partial_{x}\,\tilde{\xi}\,\partial_{t}\, \tilde{\phi}\Big)
    \;,
\end{flalign}
where $\tilde{\xi}$ and $\tilde{\phi}$ are two scalar fields.
Hence, we see that a naïve Kaluza-Klein reduction of an Abelian Chern-Simons theory on a circle yields a $\theta$-term \eqref{eq:theta_reduced} and a twist term \eqref{eq:twisted_reduced}. Despite disagreeing with the boundary-theory reduction \eqref{eq:bdy_action}, the current Kaluza-Klein-reduced theory can be made consistent with the previous results if $\partial_{x}\tilde{\xi} = 0$ and $\partial_{t}\tilde{\xi} = 2\kappa\, \partial_{t}\tilde{\phi}$.\\

A final subtle but important remark is in order. It has appeared in recent work, that in a manifold equipped with spin structure, the Chern-Simons term becomes a spin--TQFT, and the above reduction should be extended with the incorporation of a coupling to said spin structure. This is achieved by deforming the theory to harbour a $\mathbb{Z}_{2}$ index known as the \textit{Arf invariant}. We will not discuss this case here, but details can be found in Refs. \cite{karch2019arfmain,senthil2019duality,okuda2021u}, whose roots date back to the seminal work of Dijkgraaf and Witten\cite{dijkgraaf1990topological}.

\subsection{Reduced Topological Gauge Theory and Coupling to Matter}

Equation \eqref{eq:reduced} constitutes the topological gauge action in 1+1D playing a similar role to the Chern-Simons term in 2+1D. We have shown that it can be obtained as a dimensional reduction from it. We have witnessed the emergence of chiral scalar field not present in the parent theory. This scalar field encodes the information of the transverse component of the statistical gauge field, i.e. it is a local measure of the flux in the reduced dimension. Such an action is now considered as a model on its own right, which we call the \textit{chiral axion} (or $\chi-$axion) theory
\begin{equation}\label{eq:chiral_ax1}
S_{\,\chi-\text{axion}} \,[\phi_{L,R}\,,a_{\mu}] = \,\frac{1}{4\pi} \int dt\,dx\;\Big(\pm \kappa\,\partial_{t}\,\phi_{L,R}\,\partial_{x}\,\phi_{L,R} -  \phi_{L,R} \,\epsilon^{\,\mu \nu}\partial_{\,\mu}\,a_{\nu} \Big)\;.
\end{equation}
The first term $S_{\chi}[\phi_{L,R}]\,$ corresponds to a chiral boson  constraining the scalar field to move either to the right ($-$) (or to the left $+$). This implies that in momentum space we shall only sum over $k>0$ (or $k<0$). The second term $S_{\text{axion}}[\phi_{L,R}\,,a_{\mu}]\,$ is a background field (BF) contribution of the same form as a topological $\theta$-term in 1+1D. Field $\phi_{L,R}$ plays the role of a spacetime varying $\theta$-angle. More plainly, $S_{\text{axion}}[\phi_{L,R}\,,a_{\mu}]\,$ is, more accurately, the low-dimensional relative of an axion term, and $\phi_{L,R}$ is an axion. The axionic contribution alone leads to a decoupling from the statistical gauge field, but the introduction of the term $S_{\chi}[\phi_{L,R}]\,$ endowes the axion with chiral dynamics. As expected, both these contributions give a vanishing Hamiltonian, since they are first order in time derivatives, and are universal in that they do not depend on a specific matter model. The chiral axion model first appeared in Ref. \cite{aglietti1996solitons}. In that work, the authors reduce a Chern-Simons term by ``brute-force'' collapse of a spatial dimension, which leads to the appearance of the $\theta$-term. Then, they introduce the coupling to a chiral boson ``by hand''. This results in obtaining Eq. \eqref{eq:chiral_ax1} in its classical version. Here we have consistently obtained such a theory as the natural reduction of the parent theory, i.e. without the need of introducing dynamics by hand. Quantisation of the original theories is transferred to the reduced theory according to our formulation. Alternatively, we can canonically quantise the classical theory by postulating 
\begin{flalign}
	&\big[\hat{\phi}_{L,R}(x), \hat{\Pi}^{\,t}_{\phi}(x')\big] = i\hbar\,\delta\,(x-x')\;, \\
	&\big[\hat{a}_{\nu}(x), \hat{\Pi}^{\,t \,\mu}_{a}(x')\big] = i\hbar\,\delta_{\nu}^{\,\mu}\,\delta\,(x-x')\;,
\end{flalign}
where canonically conjugate momenta are defined as
\begin{flalign}
	& \hat{\Pi}^{\,t}_{\phi}(x)\equiv \frac{\partial \mathcal{L}}{\partial (\partial_{t}\phi_{L,R})} = \pm \frac{\kappa}{4\pi}\,\partial_{x}\hat{\phi}_{L,R}(x)\;, \\
	& \hat{\Pi}^{\,t \,\mu}_{a}(x)\equiv \frac{\partial \mathcal{L}}{\partial (\partial_{t} a_{\mu})} = - \frac{1}{4\pi}\,\hat{\phi}_{L,R}(x)\;.
\end{flalign}
This yields the commutation relations
\begin{flalign}
	&\big[\hat{\phi}_{L,R}(x), \hat{\phi}_{L,R}(x')\big] = \mp i\hbar \,\frac{2\pi}{\kappa}\,\text{sgn}\,(x-x')\;, \\ \label{eq:kac_moody_ex}
	&\big[\hat{a}_{x}(x), \hat{a}_{x}(x')\big] = i\hbar \,4\pi\,\partial_{x}\,\delta\,(x-x')\;,
\end{flalign}
where we have used the classical equations of motion and the identity $\partial_{x}\text{sgn}\, (x) = 2\,\delta(x)\,$, see e.g. \cite{ezawa2013quantum,fradkin2013}. This defines a $\text{U} \,(1)$ Kac-Moody algebra, so the theory is anomalous, as expected from knowledge of FQH edge states. Despite being locally defined at a given spacetime position, the chiral axion has intrinsic anomalous behaviour as it does not commute with itself at different positions. On the other hand, the statistical gauge field satisfies chiral-current-like commutation relations. These aspects become crucial when minimally coupling a chiral axion to matter. From the composite particle duality we expect an analogue of flux attachment to take place in addition to the statistical transmutation of matter fields. Now, the coupled chiral-axion-matter theory reads $\mathcal{L} = \mathcal{L}_{\chi - \text{axion}}[\phi, a] - J^{\mu}a_{\mu}$. The classical equations of motion read
\begin{flalign}
 &J^{\mu} = - \frac{1}{4\pi} \epsilon^{\mu \nu}\partial_{\nu} \phi_{L,R}\;, \\
&\epsilon^{\mu \nu} \partial_{\mu} a_{\nu} = \mp \kappa \,\big(\partial_{t} \partial_{x} \phi_{L,R} + \partial_{x} \partial_{t} \phi_{L,R} \big)\;,
\end{flalign}
from which we can extract the form of the statistical gauge field as a function of matter. For instance, in the temporal gauge, we find $a_{t} = 0$ and
\begin{equation}\label{eq:redfluxatt}
a_{x}(t,x) = \pm 4\pi \kappa\, J^{0}(t,x) + \xi(x)\;,
\end{equation}
with $\xi$ being an integration constant. This local constraint is the 1+1D version of flux attachment. The theory yields a gauge potential linear in Noether charge density, as expected. While this is fine at a classical level, at a quantum level, due to the behaviour of the Kac-Moody algebra, this Noether charge acquires anomalous commutation relations. This is at odds with conventional examples of Noether charge densities. For instance, for a non-relativistic Bose field $\hat{J}^{0} = \hat{n}=  \hat{\Psi}^{\dagger} \hat{\Psi}$, the commutation relation 
\begin{equation}\label{eq:normalCurrent}
    \big[\hat{J}^{0}(x),\hat{J}^{0}(x')\big] = 0\;,
\end{equation}
which is in contradiction with Eq.\eqref{eq:kac_moody_ex}. Hence, we face a conundrum in which one of these two options should be considered:
\begin{enumerate}
    \item The chiral axion theory should be taken as a classical theory and only quantised after effectively reducing it to a Gauss's law\footnote{ This option implicitly considered in the quantum discussion of the effective model in Ref. \cite{jackiw1990soliton}.}. The chiral axion is effectively integrated out of the theory as it is an auxiliary field, so that the effective model knows nothing about its properties. Quantum matter then dictates the commutation relations of the gauge field and there is no apparent anomalous behaviour. 

    \item The chiral axion theory should be coupled to chiral matter fields. For non-chiral matter there is an inconsistency, which requires the incorporation of left- and right-chiral sectors. The left-chiral axion couples to the left-chiral sector of matter and otherwise. In essence, the theory needs a chiral doubling.
\end{enumerate}
\noindent
Considerations $(1.)$ and $(2.)$ can be brought to agreement by assuming that each chiral axion couples to the corresponding chiral matter sector. We call this situation a doubled chiral axion ($2\chi-\text{axion}$). This means that 
\begin{equation}
    \big[\hat{J}^{0}_{L,R}(x), \hat{J}^{0}_{L,R}(x')\big] = \pm \frac{i\hbar}{4\pi \kappa^{2}}\,\partial_{x}\,\delta\,(x-x')
\end{equation}
Then, given the decomposition $\hat{J}^{0} = \hat{J}^{0}_{L} + \hat{J}^{0}_{R}\,$ and considering no cross-coupling between modes with opposite chiralities, we find that the non-chiral --- i.e. conventional --- Noether charge is non-anomalous and, thus, satisfies Eq. \eqref{eq:normalCurrent} as hoped. This is then consistent with consideration (1.). This concludes our discussion in which we have illustrated, by dimensional reduction arguments, that a gauge potential linear in Noether current density is the low-dimensional relative of flux attachment. 

\subsubsection{Statistical Transmutation on the Line}
The statistical character of a density-dependent gauge field of the form $\hat{a}_{x} \propto \hat{n}$ in one spatial dimension is known since the seminal works of Rabello \cite{rabello1996prl} and Kundu \cite{kundy99anyons}. Minimally coupling a matter field to such a statistical gauge field yields
the equivalence
\begin{equation}
    H_{\text{B}} = -\frac{\hbar^{2}}{2m} \big( \hat{D}_{x} \hat{\Psi}_{\text{B}} \big)^{\dagger} \big(\hat{D}_{x} \hat{\Psi}_{\text{B}}\big) \;\;\;\;\;\;\;\; \leftrightarrow  \;\;\;\;\;\;\;\; H_{\text{C}} = -\frac{\hbar^{2}}{2m} \big( \partial_{x} \hat{\Psi}_{\text{C}} \big)^{\dagger} \big(\partial_{x} \hat{\Psi}_{\text{C}}\big)\;\;,
\end{equation}
between a bosonic bare theory (B) and an anyonic composite theory (C), where the gauge-covariant derivative is defined as $\hat{D}_{x} = \partial_{x} -\frac{i}{\alpha} \hat{n}\,$ for the real-valued coupling $\alpha$. The passage from one theory to the other is enabled by a statistical gauge transformation of Jordan-Wigner type
\begin{equation}
    \hat{\Psi}_{\text{B}} \,(t,x) = e^{\frac{i}{\alpha} \int^{\infty}_{-\infty} dx'\,\Theta \,(x-x')\,\hat{n}\,(t,x')}\,\hat{\Psi}_{\text{C}} \,(t,x)\,,
\end{equation}
where $\Theta\,(x)$ is the Heaviside step function. Commutation relations are explicitly deformed upon removal of the gauge field, rendering the transformed matter field effectively anyonic, in close analogy to the case for the FQHE in one dimension higher \cite{ezawa2013quantum}. We start by computing the equal-time commutation rules between the Jordan-Wigner string 
\begin{equation}
    \hat{\Phi}\,(x) = \frac{1}{\alpha} \int_{-\infty}^{x}dx'\;\hat{\Psi}_{\text{B}}^{\dagger}(x') \,\hat{\Psi}_{\text{B}}(x')\;.
\end{equation}
and the bare matter field $\hat{\Psi}_{\text{B}}\,(x)$. These take the form
\begin{equation}
	\big[\hat{\Phi}\,(x')\,, \hat{\Psi}_{\text{B}}\,(x)\big] = \frac{1}{\alpha}\,\int dx'' \; \Theta\,(x' - x'')\,\big[ \hat{n}\,(x''), \hat{\Psi}_{\text{B}}\,(x) \big]= -\,\frac{1}{\alpha}\, \Theta\,(x'-x)\,\hat{\Psi}_{\text{B}}\,(x)
\end{equation}
We note that the value $\Theta\,(0)$ is to be fixed by convention, so we shall restrict our discussion to $x \neq x'$, and discuss $x=x'$ as a particular case later on.

We want to compute 
\begin{equation}
\hat{\Psi}_{\text{C}}\,(x) \,\hat{\Psi}_{\text{C}}\,(x') = e^{-i\,\hat{\Phi}\,(x)} \,\hat{\Psi}_{\text{B}}\,(x) \,e^{-i\,\hat{\Phi}\,(x')} \,\hat{\Psi}_{\text{B}}\,(x')\,.   
\end{equation}
In order to make progress, it is useful to evaluate an expression of the form
\begin{flalign}
	e^{i\,\hat{\Phi}\,(x')} \,\hat{\Psi}_{\text{B}}\,(x)\,e^{-i\,\hat{\Phi}\,(x')} &= \hat{\Psi}_{\text{B}}\,(x) + i\,\Big[\hat{\Phi}\,(x'),\hat{\Psi}_{\text{B}}\,(x)\Big] + \frac{i^{2}}{2}\Big[ \hat{\Phi}\,(x') , \Big[\hat{\Phi}\,(x'),\hat{\Psi}_{\text{B}}\,(x)\Big] \Big] + \;\dots \\
 &= e^{-\frac{i}{\alpha}\,\Theta\,(x'-x)}\,\hat{\Psi}_{\text{B}}\,(x)\,
\end{flalign}
where we have used the Baker–Campbell–Hausdorff formula 
\begin{equation}
    e^{A}B\,e^{-A} = \sum_{n} \frac{1}{n!}\,(\text{ad}_{A})^{n}  B
\end{equation}
with the linear adjoint operator defined as $\text{ad}_{A}\, B = [A,B]\,$. This allows us to write
\begin{equation}
	\hat{\Psi}_{\text{C}}\,(x) \,\hat{\Psi}_{\text{C}}\,(x') = e^{-\frac{i}{\alpha}\,\Theta\,(x'-x)} e^{-i\,\hat{\Phi}\,(x)} e^{-i\,\hat{\Phi}\,(x')} \,\hat{\Psi}_{\text{B}}\,(x)\,\hat{\Psi}_{\text{B}}\,(x')
\end{equation}
and similarly
\begin{equation}
	\hat{\Psi}_{\text{C}}\,(x') \,\hat{\Psi}_{\text{C}}\,(x) = e^{-\frac{i}{\alpha}\,\Theta\,(x-x')} e^{-i\,\hat{\Phi}\,(x')} e^{-i\,\hat{\Phi}\,(x)} \,\hat{\Psi}_{\text{B}}\,(x')\,\hat{\Psi}_{\text{B}}\,(x)\,.
\end{equation}
Commuting the bosonic and string operators among themselves, using the identity 
\begin{equation}
\Theta\,(x-x') = 1- \Theta\,(x'-x)
\end{equation}
and the definition $\text{sgn}\,(x) \equiv 2\,\Theta\,(x) -1\,$, we find
\begin{equation}
	\hat{\Psi}_{\text{C}}\,(x) \,\hat{\Psi}_{\text{C}}\,(x')  - e^{\,\frac{i}{\alpha} \,\text{sgn}\,(x-x')}\, \hat{\Psi}_{\text{C}}(x')\, \hat{\Psi}_{\text{C}}(x) = 0 
\end{equation}
Similarly, we see that
\begin{flalign}
	&\hat{\Psi}_{\text{C}}^{\dagger}\,(x) \,\hat{\Psi}_{\text{C}}^{\dagger}\,(x') - e^{\,\frac{i}{\alpha} \,\text{sgn}\,(x-x')} \,\hat{\Psi}_{\text{C}}^{\dagger}\,(x')\, \hat{\Psi}_{\text{C}}^{\dagger}\,(x) = 0 \\
	&\hat{\Psi}_{\text{C}}\,(x) \,\hat{\Psi}_{\text{C}}^{\dagger}\,(x') - e^{\,-\frac{i}{\alpha} \,\text{sgn}\,(x-x')} \,\hat{\Psi}_{\text{C}}^{\dagger}\,(x')\, \hat{\Psi}_{\text{C}}\,(x) = \delta\,(x-x')
\end{flalign}
are satisfied. We now note for the convention $\text{sgn}\,(0) = 0$, we recover bosonic commutation relations at $x=x'$, provided the string contributions cancel each other out and we are left with the statistics of the bare fields. This can be understood of a particular instance of a composite particle duality \cite{valenti2024dual,valenti2022topological}, which is briefly reviewed for completeness in Appendix \ref{sec:compo}.
\section{Conclusions}
Building on previous works \cite{aglietti1996anyons,jackiw2000reduction} and based on the intuition gained from \cite{valenti2024dual,valenti2022topological}, we have found a reduced theory \eqref{eq:chiral_ax1} for an Abelian Chern-Simons term and the corresponding analogue of a flux attachment law \eqref{eq:redfluxatt} when coupled to non-relativistic matter. It is clear that such a remnant can not be interpreted, strictly speaking, as attaching flux to charges, but rather as ``gauge dressing'' the corresponding matter field. Removing such a statistical gauge field comes at the expense of changing the quantum identity of the matter it is coupled to, as it happens in the more familiar parent theory describing the long-wavelength behaviour of FQH fluids. 

The reduced theory has two contributions, an axion-like term and a chiral term for said axion. Our reduction agrees with the proposed theory in \cite{aglietti1996anyons}, but appears naturally without the need of complementing the reduction by introducing terms \textit{ad hoc}. Two important caveats are, firstly, that the concept of ``dimensional reduction'' is vaguely defined and it is therefore sensitive to the techniques used. Secondly, there is an intrinsic anomalous behaviour in the quantisation of the reduced theory not highlighted in previous works that probably deserves adequate care beyond the scope of the discussion in our work.

\section*{Acknowledgments}
We thank N. Rougerie for stimulating discussions. G. V-R. acknowledges financial support from the Naquidis Center for Quantum Technologies.





\bibliography{SciPost_Example_BiBTeX_File.bib}


\appendix

\newpage
\section{The Chern-Simons term as a total divergence}\label{sec:divergence}

Here we provide a geometric view on the Chern-Simons term. We may expect the Chern-Simons theory to be written as a total derivative, and hence integrated to the boundary via Stokes' Theorem. This is precisely how the Chern-Simons term is introduced sometimes, as descending from the $\theta$-term in 3+1D electrodynamics. Let us work in form language and consider the so-called Clebsch parametrisation or Darboux theorem stating that any 1-form $a \equiv a_{i} \,\text{d}x^{\,i}$ can be written as
\begin{equation}
	a = \text{d}\theta + \alpha \text{d} \beta\;.
\end{equation}
Thus, the Abelian Chern-Simons Lagrangian density is a total divergence
\begin{equation}
	a\text{d}a = \text{d}\theta \text{d} \alpha \text{d} \beta = \text{d}\,(\theta \text{d} \alpha \text{d} \beta) = - \text{d}\,(\text{d}\theta \alpha \text{d}\beta) = \text{d}\,(\text{d}\theta \text{d}\alpha \beta)
\end{equation}
or, in components
\begin{equation}
	\epsilon^{\,\mu \nu \lambda} a_{\mu} \partial_{\nu} a_{\lambda} = \epsilon^{\,\mu \nu \lambda} \partial_{\mu} \theta \partial_{\nu} \alpha \partial_{\lambda} \beta = \partial_{\mu} (\epsilon^{\,\mu \nu \lambda} \theta \partial_{\nu} \alpha \partial_{\lambda} \beta)\;.
\end{equation}	
This is a total derivative and it can be integrated to the boundary by virtue of the divergence theorem. Notice that the Clebsch parameters $\theta(\mathbf{x})$, $\alpha(\mathbf{x})$ and $\beta(\mathbf{x})$ are scalar fields. A possible identification is $a_{\mu}=\theta \partial_{\mu}\beta$ and $\alpha = \phi$. This is then an Aharonov-Bohm twist term, which can be re-expressed as a 1+1D axion term $\phi \epsilon^{\mu \nu} \partial_{\mu} a_{\nu}\,$. A similar argument can be obtained by means of Helmholtz-Hodge decomposition.\\

The prescription in terms of Clebsch parameters is ubiquitous in the study of magneto-hydrodynamics where the Chern-Simons term in Euclidean space is thought of as a \textit{magnetic helicity} of the form 
\begin{equation}
	\Gamma = \int \text{d}^{3}\mathbf{x}\; \bm{a}\cdot\bm{b}\;,
\end{equation}
which measures the \textit{linkage} of magnetic flux lines, and where $\bm{b} = \bm{\nabla} \times \bm{a}$. The term \textit{helicity}, also known as \textit{Hopf invariant}, comes from fluid dynamics where the analogous term is the \textit{kinetic helicity} 
\begin{equation}
	\mathcal{H} = \int_{\mathbb{R}^{3}} v\wedge \text{d}v = \int \text{d}^{3}\mathbf{x}\; \bm{v}\cdot\bm{\omega}\;,
\end{equation}
where $\bm{v}$ is the local fluid velocity and $\bm{\omega} = \bm{\nabla} \times \bm{v}$ is the vorticity \cite{jackiw2004perfect}.

\section{The Composite Particle Duality in a Nutshell}\label{sec:compo}
The minimal coupling of charged matter to a $\text{U} \,(1)$ \textit{statistical gauge field} $a_{\mu}$ in $\text{D}=d+1$ spacetime dimensions is equivalent to the formation of electric-magnetic entities identified as gauge-charge composites. In some instances, the latter may be regarded as anyons. This correspondence is summarised as
\begin{equation}\label{eq:comp_part_duality}
\mathcal{H}_{\text{\,B}} = \sum_{i=1}^{N} \frac{\bm{\pi}_{i}^{\,2}}{2m} + \mathcal{H}_{\text{int}}\;\; \longleftrightarrow \;\; \tilde{\mathcal{H}}_{\text{\,C}} = \sum_{i=1}^{N}\frac{\tilde{\bm{p}}_{i}^{\,2}}{2m} + \tilde{\mathcal{H}}_{\text{int}}\;\;,
\end{equation}
where $\bm{\pi}_{i} = \bm{p}_{i} - \bm{a} \,(t,\mathbf{x}_{i})\,$, $N$ is the number of particles, and interactions are short-ranged. We postulate that the statistical gauge potential is a topologically non-trivial pure gauge configuration
\begin{equation}\label{eq:pure_gauge2}
	\bm{a}(t,\mathbf{x}_{i})  = \bm{\nabla}_{\mathbf{x}_{i}} \Phi\,(t,\mathbf{x}_{1},\dots,\mathbf{x}_{N})\;,
\end{equation}	
where $\bm{a}\,(t,\mathbf{x}_{i}) \equiv \bm{a}\,(t,\mathbf{x}_{i}\,;\,\mathbf{x}_{1},\dots,\mathbf{x}_{N}) $ refers to the gauge potential being evaluated at the location of particle $\mathbf{x}_{i}$, although it might be a function of the position of all the particles in the system and, as we will find later, also of matter density $\lvert \Psi_{\text{B}}\,(t,\mathbf{x}_{1},\dots,\mathbf{x}_{N})\rvert^{2}\,$. We identify a many-body Hamiltonian in the \textit{bare} (B) basis, and another corresponding to the \textit{composite} (C) basis. Both sides of the duality are related by a large gauge transformation which removes/introduces a minimally coupled statistical gauge field. In doing so, it connects homotopically distinct states with the same physical properties. This transformation corresponds to the naïve generalised continuum version of the well-known Jordan-Wigner transformation which, in second-quantised language, reads
\begin{equation}\label{eq:jw_transf}
\hat{\Psi}_{\text{C}}\,(t,\mathbf{x}; \Gamma_{\mathbf{x}}) =	\hat{\mathcal{W}}^{\,\dagger} (t,\mathbf{x};\, \Gamma_{\mathbf{x}}) \,\hat{\Psi}_{\text{B}}\,(t,\mathbf{x}) \;.
\end{equation}
The operator 
\begin{equation}
    \hat{\mathcal{W}}\,(t,\mathbf{x};\Gamma_{\mathbf{x}})= e^{\,\frac{i}{\hbar}\,\hat{\Phi}\,(t,\mathbf{x};\, \Gamma_{\mathbf{x}})}
\end{equation} 
is identified as a \textit{disorder} operator, $\hat{\Phi}$ is a $(\text{D}-1)-$dimensional \textit{Jordan-Wigner brane}, and $\Gamma_{\mathbf{x}}$ is a reference contour centred at $\mathbf{x}$ in the sense of Ref. \cite{marino2017quantum} but physically identified as a Dirac string or open 't Hooft line. This is meant to generalise the concept of a Jordan-Wigner string to higher dimensions. For low dimensions the \textit{brane}, and thus the composite operator, are local, i.e. they can be defined at a given point in space. However, for $\text{D}\ge 3$ this is not the case and these objects become intrinsically non-local, a feature captured by $\Gamma_{\mathbf{x}}$.\\

Correspondence in Eq. \eqref{eq:jw_transf} can be understood as an operator identity valid regardless of the underlying Hamiltonian. Considering the bare species $\hat{\Psi}_{\text{B}}$ to be a bosonic (fermionic) field, and hence satisfying ordinary equal-time (anti)commutation  relations, then, $\hat{\Psi}_{\text{C}}$ constitutes a composite field obeying generalised commutation relations, except for at the point $\mathbf{x} = \mathbf{x'}$, where relations reduce to those of bare species due to cancellation of branes. The density operator is $\hat{n} \,(t,\mathbf{x}) = \hat{\Psi}_{\text{B}}^{\dagger} (t,\mathbf{x})\,\hat{\Psi}_{\text{B}}(t,\mathbf{x}) = \hat{\Psi}_{\text{C}}^{\dagger}(t,\mathbf{x};\Gamma_{\mathbf{x}}) \,\hat{\Psi}_{\text{C}}(t,\mathbf{x};\Gamma_{\mathbf{x}})\,$, so all possible local interaction terms which are functions of the density are identical on both sides of the duality. The correspondence also dictates the functional form of the statistical gauge field. In $d\leq 3$ and considering a temporal gauge, they result from solving
\begin{flalign}
&\text{1+1D}\;:\;\;\;\;\;\;\;\;\;\;\;\;\;\;\;\;\;\;\;\;\;\;\;\;\;\;\;\;\;\;\;\;\;\;\;\; \frac{2 \pi}{\kappa}\, \hat{n}\,(t,x) = \hat{a}_{x} \,(t,x) &\\\
&\text{2+1D}\;:\;\;\;\;\;\;\;\;\;\;\;\;\;\;\;\;\;\;\;\;\;\;\;\;\;\;\;\;\;\;\;\;\;\;\;\; \frac{2 \pi}{\kappa}\, \hat{n}\,(t,\mathbf{x}) = \bm{\nabla} \times \hat{\bm{a}} \,(t,\mathbf{x})&\\ 
&\text{3+1D}\;:\;\;\;\;\;\;\;\;\;\;\;\;\;\;\;\;\;\;\;\;\;\;\;\;\;\;\;\;\;\;\;\;\;\;\;\; \frac{2 \pi}{\kappa}\,\hat{n} \,(t,\mathbf{x})= \bm{\nabla} \cdot \bm{\nabla}\times \hat{\bm{a}}  \,(t,\mathbf{x})\;,
\end{flalign}
where the matter current $\partial_{t}\hat{n}\,(t,\mathbf{x}) + \bm{\nabla}\cdot \hat{\bm{J}}\,(t,\mathbf{x}) = 0$ is conserved. We can make the following general observations: 
\begin{itemize}
    \item The composite particle duality naturally generalises those of conventional flux attachment and statistical transmutation. It survives in any dimension, its origin is geometric, it is physically enforced by topological terms, and satisfies an \textit{order-disorder} operator structure.
    \item Bose-Fermi correspondences can be seen as statistical transmutation. Hence, they can be probed in experiments by gauge-coupling quantum matter and tuning the coupling constant of the statistical gauge field. This will correspond to a physical interpolation between faces of the duality or, equivalently, changing the statistical parameter. We also expect the presence of exotic gauge-charge composite quasiparticles.
\end{itemize}
A crucial implication of our formalism is that Eq. \eqref{eq:comp_part_duality} not only signals an identification between theories, but also a physical process of gauge-dressing and transmutation of matter fields. This means that some ``conventional'' interaction between matter fields can be interpreted as a ``statistical'' interaction. In other words, statistical gauge fields can be just a relabelling of certain conventional interactions in quantum matter. This provides new meaning to the \textit{emergence} of gauge fields in quantum many-body systems.

\end{document}